\documentclass{PoS}
\usepackage{wrapfig}
\usepackage{epsfig}

\title{Scaling properties in deep inelastic scattering}

\ShortTitle{Scaling properties in DIS}

\author{\speaker{Christophe Royon}\\
        CEA/IRFU/Service de physique des particules, CEA/Saclay, 91191 
Gif-sur-Yvette cedex, France\\
        E-mail: \email{christophe.royon@cea.fr}}

\author{Robert Peschanski\\
        CEA/Service de physique th\'eorique, CEA/Saclay, 91191 
Gif-sur-Yvette cedex, France\\
        E-mail: \email{robert.peschanski@cea.fr}}

\abstract{We study the scaling properties in deep inelastic scattering using the
most recent combined structure function data $F_2$ from the H1 and ZEUS
collaborations. We also perform a direct fit to the $F_2$ data inspired by the
scaling properties. Our analysis favours the QCD saturation mechanism from the
Balitski Kovchegiv equation wuth running coupling.
}

\FullConference{XVIII International Workshop on Deep-Inelastic Scattering and Related Subjects, DIS 2010\\
		April 19-23, 2010\\
		Firenze, Italy}

\begin{document}

\section{Scalings in deep inelastic scattering}
Geometric scaling~\cite{Stasto:2000er,us} is a remarkable empirical 
property found originally using the 
data on high energy deep inelastic scattering (DIS) $i.e.$
virtual photon-proton cross-sections. 
One can  represent with 
reasonable 
accuracy the cross section $\sigma^{\gamma^*p}$ by the formula
$\sigma^{\gamma^*p}(Y,Q)=\sigma^{\gamma^*}(\tau)\  ,$
where $Q$ is the virtuality of the photon,
$Y$ the total rapidity in the ${\gamma^*}$-proton system and 
\begin{equation}
\tau = \log Q^2-\log Q_s(Y) =  \log Q^2-\lambda 
Y\  ,
\label{tau}
\end{equation}
is the scaling variable. A fit to the DIS data measured leads to a value of
$\lambda \sim .3$, which confirms the value found within
the Golec-Biernat and W\"usthoff model \cite{Golec-Biernat:1998js} 
where 
geometric scaling was explicitely used for the parametrization.

The scaling using the variable $\tau$ defined in Formula~\ref{tau} is
directly related to the concept of saturation,  
the behaviour of perturbative 
QCD amplitudes when the density of partons becomes high enough. 
There were many theoretical arguments  to 
infer that in a domain in $Y$ and  $Q^2$ where saturation effects set in, geometric 
scaling 
is  expected to occur. Within this framework, the function  $Q_s (Y)$ can be 
called the 
saturation scale, since it  determines the approximate upper bound of the 
saturation domain.

This type of geometric scaling is motivated by asymptotic properties of QCD 
evolution equations with rapidity. Using the nonlinear 
Balitsky-Kovchegov 
(BK) equation~\cite{Balitsky:1995ub} which represents the ``mean-field'' 
approximation of high energy 
(or high density) QCD, geometric scaling could be derived
from its asymptotic solutions~\cite{Munier:2003vc}.
This equation is supposed to capture some essential 
features of 
saturation effects. Considering the BK equation with $fixed$ 
coupling constant leads asymptotically to the original geometric scaling of 
Formula~\ref{tau}. Considering a $running$  coupling leads to the following
scaling
\begin{equation}
\tau = \log Q^2-\log Q_s(Y) =  \log Q^2-\lambda 
\sqrt Y\  ,
\label{taurunI}
\end{equation}

Recently~\cite{gb}, it was noticed that the scaling 
solution~\ref{taurunI} of the BK equation with running coupling is only 
approximate and not unique. Another equivalent approximation leads to a 
different scaling variable, namely 
\begin{equation}
\tau =   \log Q^2-\lambda ~ \frac{Y}{\log Q^2}
\  .
\label{taurunII}
\end{equation}

The effect of QCD 
fluctuations was examined in Ref.~\cite{Hatta:2006hs} in the fixed 
coupling scheme 
and gives rise to a new ``diffusive scaling'', the scaling variable being
\begin{equation}
\tau =   \frac{\log Q^2-\lambda {Y}}{\sqrt{Y}}
\  .
\label{diffuse}
\end{equation}

\begin{figure}
\begin{center}
\epsfig{file=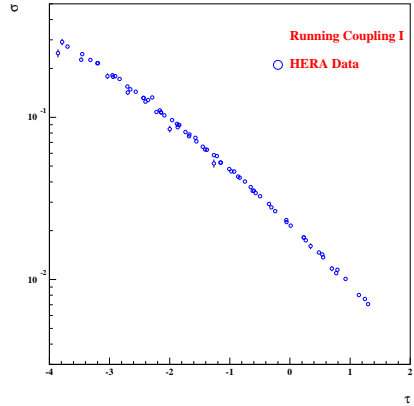,width=6.cm} 
\caption{\label{scaling}
DIS cross section data as a function of the scaling variable for Running
Coupling I.}
\end{center}
\end{figure}

The aim of this study from a theoretical point-of-view is to test and compare the 
different scaling behaviors, arising from different versions of 
QCD evolution, using the most recent precise data available from HERA resulting
from a combination of the H1 and ZEUS $F_2$ measurements~\cite{combine}.
We study the quality of the description of these combined data set using 
the four kinds of scaling and 
refer to them in the following as ``Fixed Coupling'' for the variable~\ref{tau},
``Running Coupling I'' and ``Running Coupling II'' for the 
variables \ref{taurunI} and \ref{taurunII} respectively, and to  ``Diffusive 
Scaling'' for \ref{diffuse}.

\section{The Quality Factor}
In order to compare the quality of the different scalings and to check if the
DIS cross sections $\sim F_2 / Q^2$ depend mainly on the $\tau$ variable or not,
it is useful to introduce the concept of Quality Factor~\cite{qf} (QF)
while the explicit form of the $\tau$ dependence is not known. After normalising
the data sets $v_i=\log(\sigma_i)$ and scalings $u_i=\tau_i(\lambda)$ between
0 and 1, and ordering the scalings in $u_i$, we define QF
\begin{eqnarray}
QF(\lambda) = \left[ \sum_{i} \frac{(v_i-v_{i-1})^2 }
{(u_i-u_{i-1})^2+\epsilon^2} \right]^{-1}.
\nonumber
\end{eqnarray}
$\epsilon$ is needed in the case that two data points have the same scaling,
namely when they have the same $x$ and $Q^2$, and we take $\epsilon$=0.01.
The method is to fit the value of $\lambda$ to maximize QF.

\section{Scaling tests in DIS}
As we mentioned, we use the very precise data sets combining the H1 and ZEUS
measurements of the proton structure function $F_2$~\cite{combine}. To remain in
the region where perturbative QCD is applicable and to avoid the region where
valence quarks dominate, we choose to restrict ourselves to data points with
$4 \le Q^2 \le 150$ GeV$^2$ and $x \le 10^{-2}$. In addition, in order to avoid
the high $y$ region where $F_L$ is large, we add an additional cut on data
on $y \le 0.6$. After all cuts, we are left with 117 data points.

The values of the $\lambda$ parameters and the QF are given in Table I for the
different scaling considered in this analysis. While Fixed Coupling, Running
Coupling I and II lead to approximately the sames value of QF, Diffusive Scaling
is clearly disfavoured. As an example, the scaling plot showing all combined DIS
cross section data as a function of the $\tau$ variable for Running Coupling I
is given in Fig.~I to show the quality of scaling. In addition, it is worth
noticing that adding additional variables such as $Q_0$ or a shift in rapidity
$Y_0$ does not improve the scaling quality.

\begin{table}
\begin{center}
\begin{tabular}{|c||c|c|} \hline
scaling& parameter & 1/QF  \\
\hline\hline
Fixed Coupling & $\lambda=0.31$ & 150.2 \\
Running Coupling I &  $\lambda=1.61$ & 137.9\\
Running Coupling II & $\lambda=2.76$ & 124.3\\
Diffusive Scaling & $\lambda=0.31$ & 210.7 \\\hline
\end{tabular}
\caption{Values of $1/QF$ and of the $\lambda$ parameter for the four different
kinds of scaling considered} 
\end{center}
\end{table}

\begin{figure}[t]
\hfill
\begin{minipage}[t]{.45\textwidth}

\epsfig{file=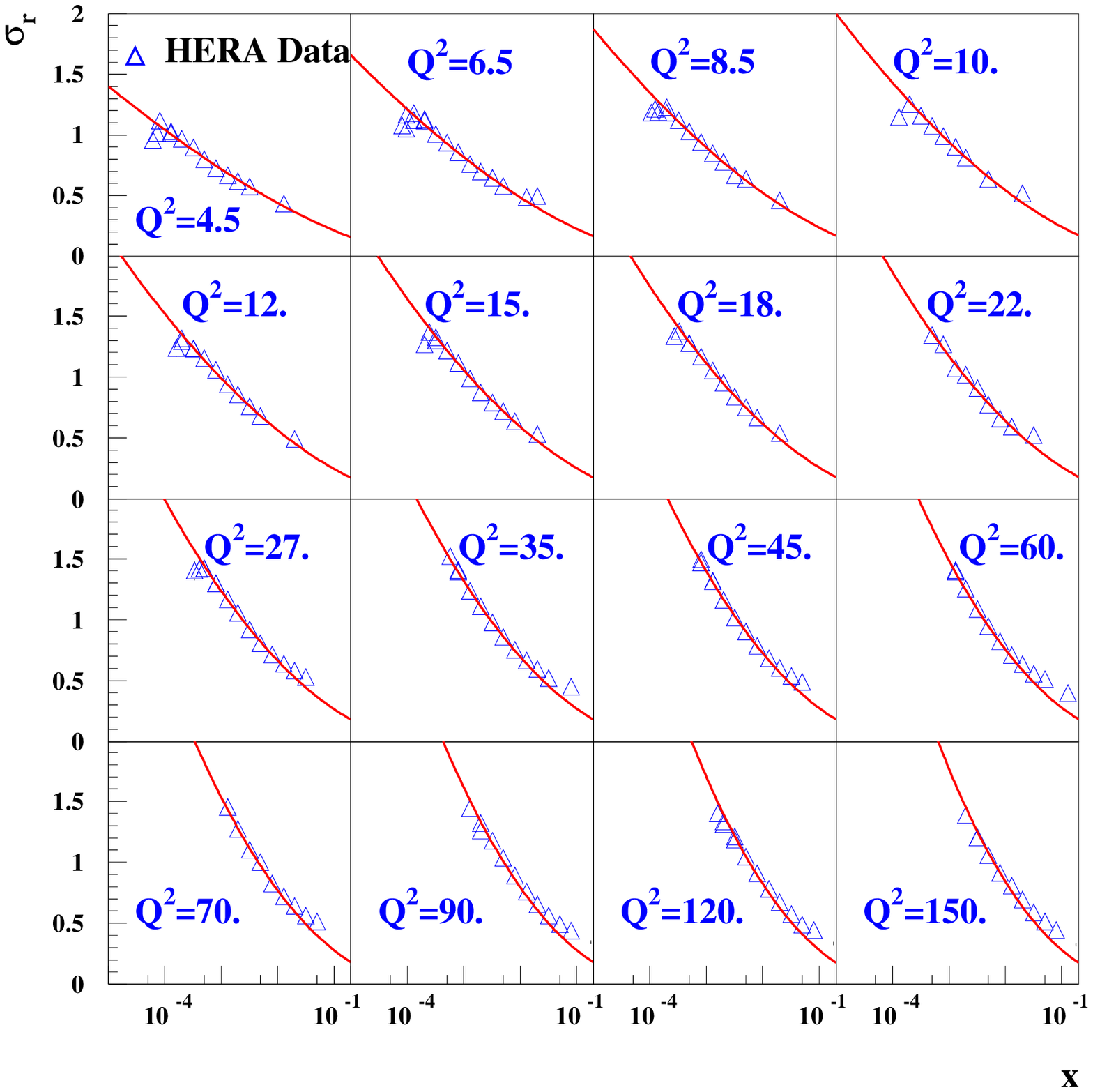,width=6.cm} 
\caption{Result of the Running Coupling I fit to the combined $F_2$ data set from H1 and
ZEUS. We note a fair description of data for $x\le 10^{-2}$ and $y \le 0.6$.}
\label{F2_QF_1}

\end{minipage}
\hfill
\begin{minipage}[t]{.45\textwidth}

\epsfig{file=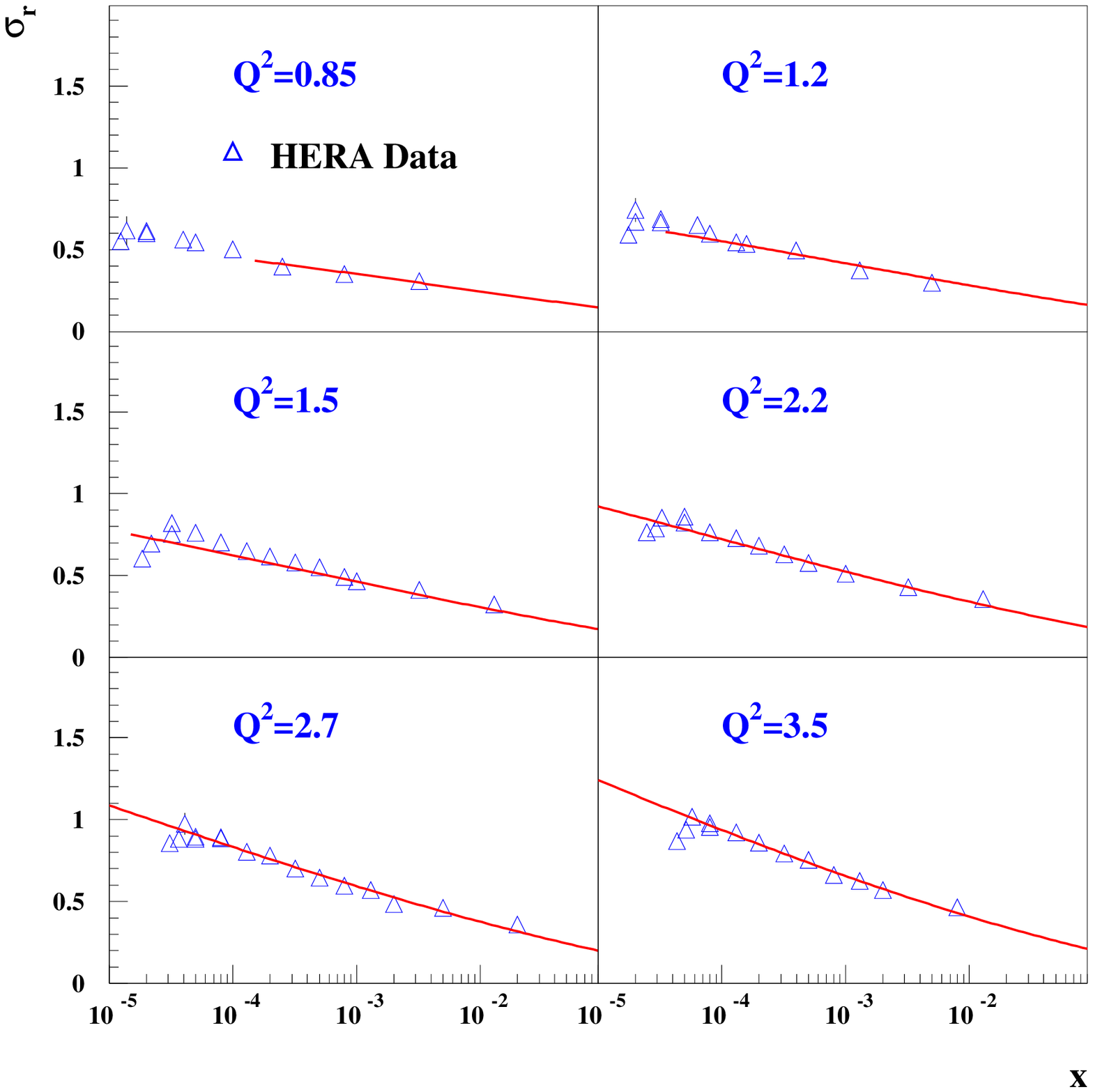,width=6.cm} 
\caption{Extrapolation of the Running Coupling I fit to the combined $F_2$ data set from H1 and
ZEUS at low $Q^2$ values.}
\label{DVCS}

\end{minipage}
\hfill
\end{figure}

\section{Fits to HERA data}
In this section, we describe a fit to the combined HERA data motivated by the
success of the data description using the Running Coupling I scaling variable.
In the fit, we will use all data above $Q^2=4 GeV^2$ since the fitting formula
that we develop is valid only in the dilute regime, and saturation is supposed
to occur at very low $Q^2$ at HERA. The following formula, 
deduced from the dipole amplitude with saturation
including the asymptotic expression of the Airy function which is
the solution of the Balistsky-Kovchegov equation, is used to fit the data
\begin{eqnarray}
\tau &=& \log (\frac{Q^2}{Q_0^2}) - \lambda 
\sqrt{\log(\frac{1}{x}) - Y_0}
\nonumber \\
\sigma &=& N \exp(-\alpha \tau)  
\exp \left( \frac{- \beta \tau^{3/2}}
{(\log{1/x} - Y_0)^{1/4}} \right) 
\nonumber
\end{eqnarray}
where the different parameters used in the fit are $\lambda$, $\alpha$, $\beta$, $Q_0$,
$Y_0$ and $N$. We notice that this formula shows only a moderate scaling
violation introduced by the $(\log{1/x} -
Y_0)^{1/4}$ term predicted by the dipole model and we perform 
the fits with and without this term.

The fit results and the parameter values are given in Table II and Fig.~II. 
We note that the fit $\chi^2$ is close to 1.2 per dof and is similar with or
without the scaling violation term. The fit cannot describe the reduction of the
reduced cross section at high $y$ due to the large values of $F_L$. In Fig.~III,
we also show the fit extrapolation at lower $Q^2$ which leads to a fair
description of data. Going to lower values of $Q^2$ will require a
parametrisation valid in the saturated region whereas our formula is only valid
in the dilute regime.

In addition, we also attempted to perform a similar fit inspired by Fixed
Coupling or Running Coupling II, but they lead to a worse description of data
($\chi^2=$156.4 and 190.4 respectively).

\begin{table}
\begin{center}
\begin{tabular}{|c||c|c|} \hline
Parameter& Fit I & Fit II  \\
\hline\hline
$\lambda$ & 1.54 $\pm$ 0.02 & 1.54 $\pm$ 0.02 \\
$\alpha$ & 0.34 $\pm$ 0.01 & 0.18 $\pm$ 0.01\\
$\beta$ & 0.24 $\pm$ 0.01 &0.18 $\pm$ 0.01 \\
$Q_0$ & 0.079 $\pm$ 0.01 & 0.064 $\pm$ 0.01\\
$Y_0$ & -1.46 $\pm$ 0.02 & 0.50 $\pm$ 0.02\\
$N$ & 0.51  $\pm$ 0,01 & 0.72  $\pm$ 0,01 \\ \hline
$\chi^2$ &  130.1 & 119.0
 \\\hline
\end{tabular}
\caption{Value of the parameters of the Running Coupling I inspired fit to the
combined DIS cross section 117 data points from HERA. Fit I (resp. Fit II) is performed
with (resp. without) the scaling violation term.}
\end{center}
\end{table}

\section{Conclusion}
In this paper, we presented a new study of scaling properties in DIS using the
most recent combined $F_2$ data from H1 and ZEUS. The new precise data set shows
precise scaling using either Fixed Coupling, Running Coupling I or II, while
Diffusive Scaling is disfavoured. A direct fit inspired by Running Coupling I
leads to a good description of data in the dilute regime.

\end{document}